\documentclass[twocolumn,showpacs]{revtex4}

\usepackage{amsmath}

\usepackage{graphicx}

\begin{document}

\title{Direct Observation of Condon Domains in Silver by Hall Probes}
\author{R.B.G.~Kramer$^{a}$, V.S.~Egorov$^{a,b}$,
V.A.~Gasparov$^{c}$, A.G.M.~Jansen$^{d}$, and W.~Joss$^{a}$}
\affiliation{$^{a}$Grenoble High Magnetic Field Laboratory,
Max-Planck-Institut f\"{u}r Festk\"{o}rperforschung, B.P. 166X, F-38042 Grenoble Cedex 9, France \\
$^{b}$Russian Research Center Kurchatov Institute, 123182 Moscow, Russia \\
$^{c}$Institute of Solid State Physics, Russian Academy of Sciences, 142432 Chernogolovka, Russia \\
$^{d}$Service de Physique Statistique, Magn\'{e}tisme, et
Supraconductivit\'{e}, D\'{e}partement de Recherche Fondamentale sur
la Mati\`{e}re Condens\'{e}e, CEA-Grenoble, F-38054 Grenoble Cedex
9, France}

\begin{abstract}
Using a set of micro Hall probes for the detection of the local
induction, the inhomogeneous Condon domain structure has been
directly observed at the surface of a pure silver single crystal
under strong Landau quantization in magnetic fields up to 10~T. The
inhomogeneous induction occurs in the theoretically predicted part
of the $H-T$~Condon domain phase diagram. Information about size,
shape and orientation of the domains is obtained by analyzing Hall
probes placed along and across the long sample axis and by tilting
the sample. On a beryllium surface the induction inhomogeneity is
almost absent although the expected induction splitting here is at
least ten times higher than in silver.
\end{abstract}
\date{\today}
\maketitle

Periodic formation and disappearance of a phase with diamagnetic and
paramagnetic domains was predicted by Condon~\cite{CONDON} to occur
in a normally nonmagnetic metal in strong magnetic fields ($H$) at
low temperatures ($T$). The domains arise under the condition
$\chi=\mu_{0}\partial M/\partial B> 1$, where $M$ is the oscillating
magnetization of the electrons due to Landau quantization and
$B=\mu_{0}(H+M)$ the total induction inside the sample. In this case
$\mu_0\partial H/\partial B=1-\chi<0$ which implies
thermodynamically unstable sections and the multivaluedness of the
induction $B(H)$ within some part of each dHvA oscillation
period~\cite{SHOEN,SOLTBaines}. For a long rod-like sample oriented
along $\mathbf{H}$, the instability is avoided by a discontinuous
jump $\delta B = B_2 - B_1$ between two stable values $B_{1}$ and
$B_{2}$ at a given critical field $H_{c}$. For a plate-like sample
perpendicular to $\mathbf{H}$, the boundary condition $B=\mu_{0}H$
for a uniformly magnetized state leads to domain formation with
alternating regions of diamagnetic and paramagnetic magnetization
for $H$ in the range $B_{1}<\mu_{0}H<B_{2}$~\cite{CONDON}. The
proportion of the domains varies with $H$ so that
$\overline{B}=\mu_{0}H$ is fulfilled as an average over the
sample~\cite{CONWAL,SOLTBaines}. The $H-B$ diagram is similar to the
$p-V$ diagram of a van der Waals gas, only with more than one
discontinuity interval $\delta B$, situated periodically on the $B$
axis.

The existence of domains has been firstly discovered by Condon and
Walstedt on a single crystal of silver~\cite{CONWAL}. The domains
were revealed by a periodic splitting of the NMR line, corresponding
to a local field difference of about 12~G between the paramagnetic
and diamagnetic regions in fields of about 9~T. This pioneering
result remained the only reference work in the next decades. New
experimental possibilities appeared with the development of muon
spin rotation. Condon domains were observed in beryllium, white tin,
aluminum, lead, and indium~\cite{SOLT,SOLTEGOROV}. By now, no doubt,
Condon domains are expected to appear in pure single crystals of all
metals. Thermodynamic aspects of the Condon domain phase transition
have been recently treated theoretically~\cite{GOR98}. While the
state of art in this field has been recently
reviewed~\cite{GORDONVagner}, some important questions however,
concerning domain size and topology, domain wall energy, and pinning
properties can only be solved with a detailed knowledge of the
domain structure.

The state with Condon domains can be considered as physically
similar to the intermediate state of type I superconductors, where
superconducting and normal regions form in an applied magnetic
field. Therefore, domain structures resulting of such different
phenomena as superconductivity and dHvA effect may be rather
similar. Unfortunately, the magnetic contrast, that is the ratio of
$\delta B$ to $B_{2}$, is not more than 0.1~$\%$ for Condon domains
(compared to 100$\%$ for the intermediate state). Besides, the
magnetic field itself is here hundred times higher. Thus, methods
like magnetic decoration or magneto-optical detection used for
intermediate state imaging~\cite{LIVINGSTONE} can not be used for
Condon domains.

In this Letter we present the first experimental results for direct
observation of Condon domain structures in silver by a system of ten
micro Hall probes being close to the single crystal surface. In the
homogeneous state, without domains, all probes show the same dHvA
signal $B(H)$, i. e. all $B_{i}=B(H)$, where $i=1,2...10$ are the
Hall probe numbers. In the domain state, the Hall voltages differ
between the different probes in the paramagnetic part of the dHvA
period. This implies an inhomogeneous magnetic field distribution
due to Condon domains at the sample surface. In our measurements the
surface of the crystal was either normal to $\mathbf{H}$ direction
or slightly tilted (13$^{\circ}$). By comparing the data of
neighboring Hall probes, new information about the domain structure
has been extracted.

\begin{figure}[!b]
     \begin{center}
   \leavevmode
       \includegraphics[width=0.6\linewidth]{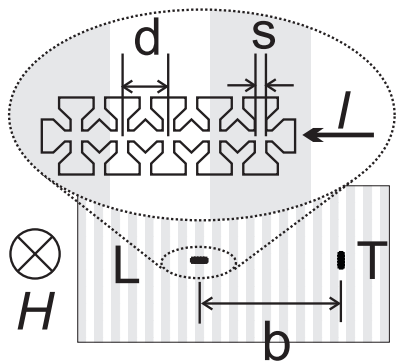}
       \caption{Experimental configuration with the
       longitudinal (L) and transverse (T) arrays of Hall probes at distance
       $b =1$~mm ($s^{2}=10 \times 10~\mu$m$^{2}$; $d=40~\mu$m).}
    \label{f-1}
\end{center}
\end{figure}\begin{figure}[!b]
     \begin{center}
    \leavevmode
       \includegraphics[width=0.95\linewidth]{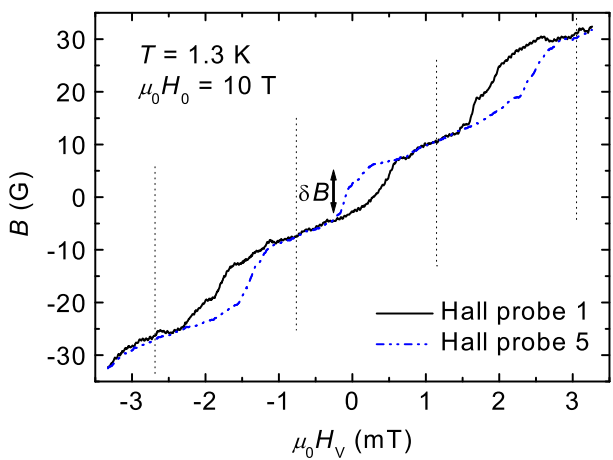}
       \caption{$B(H)$ trace for silver single
       crystal with $H \parallel [100]$ showing the splitting $\delta B$ of $B_{1}$ and
       $B_{5}$ of the L-array for three dHvA periods separated by dotted lines. The calibration of $B$ is
       with respect to the offset field $\mu_0 H_{\mathrm{0}}=10$~T.}
     \label{f-2}
     \end{center}
\end{figure}
Fig.~\ref{f-1} shows the Hall probe set-up made of a 1~$\mu$m thick
Si doped GaAs layer sandwiched between two 10~nm thick undoped GaAs
layers. Two arrays of five Hall probes (10$\times$10~$\mu$m$^2$ at
40~$\mu$m distance) are placed at a distance of $b=1$~mm. One array,
L, is oriented along the long axis of the sample; the other, T,
transverse to this axis. A DC Hall current of 100~$\mu$A was applied
in series to all five Hall probes of an array. The Hall voltages
were read out simultaneously by 5 Keithley multimeters; the arrays L
and T were measured one after another. Due to the 3D conducting
layer the $V_i(B)$ characteristics of the Hall probes were in good
approximation linear up to 10~T even at 1.3~K. The correct
calibration of the Hall probes was tested at temperatures between
4.2-3.6~K where all Hall probes showed exactly the same dHvA
oscillations of the homogenous silver sample. The detection limit of
the Hall probes was smaller than 1~G. A high homogeneity (better
than 10 ppm in 1~cm$^3$) 10~T superconducting magnet was used to set
a fixed offset magnetic field $H_0$. The slowly varying superimposed
field $H_V$ ($\pm 15$~mT) was made by a watercooled resistive coil.
Thus the total applied magnetic field was $H=H_0+H_V$.

The measurements were performed on a high quality silver single
crystal of $2.4\times 1.6\times 1.0$~mm$^{3}$. The largest surface
of the sample was normal to the $[100]$-axis of the crystal. The
sample was prepared in the same way as in experiments on radio
frequency size effect and time of flight effect (see references
in~\cite{GASPAROV}). The very good quality of the sample results in
a very low Dingle temperature, which was estimated from our
measurements to be about $T_{D}=0.2$~K. The sample was annealed in
O$_{2}$ (10$^{-2}$~Pa) at 750$^{\circ}$C during 10 hours. It has a
residual resistance ratio $RRR=
R_{300K}/R_{4.2K}=1.6\times$10$^{4}$, measured by the contactless
Zernov-Sharvin method~\cite{ZERNOVSHARVIN}. For a mirror-like
surface, the crystal was slightly repolished by 0.1~$\mu$m diamond
paste after annealing. The surface before polishing had a roughness
of about 20-30~$\mu$m and no induction splitting due to Condon
domains could be observed. The sample was glued by narrow strips of
cigarette-paper to the set-up frame to fix the crystal on the Hall
probes in order to avoid damage or strain of the single crystal upon
cooling down. Moreover, the sample was slightly pressed by a cotton
tampon to hold it reliably in high magnetic field.

\begin{figure}[!h]
     \begin{center}
   \leavevmode
       \includegraphics[width=0.95\linewidth]{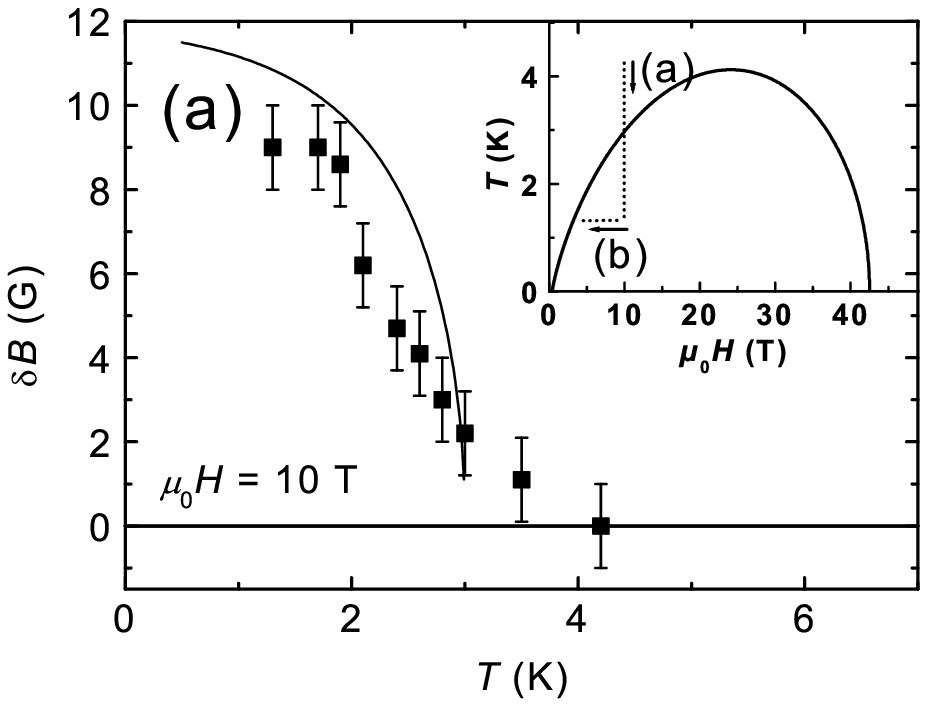}
       \includegraphics[width=0.95\linewidth]{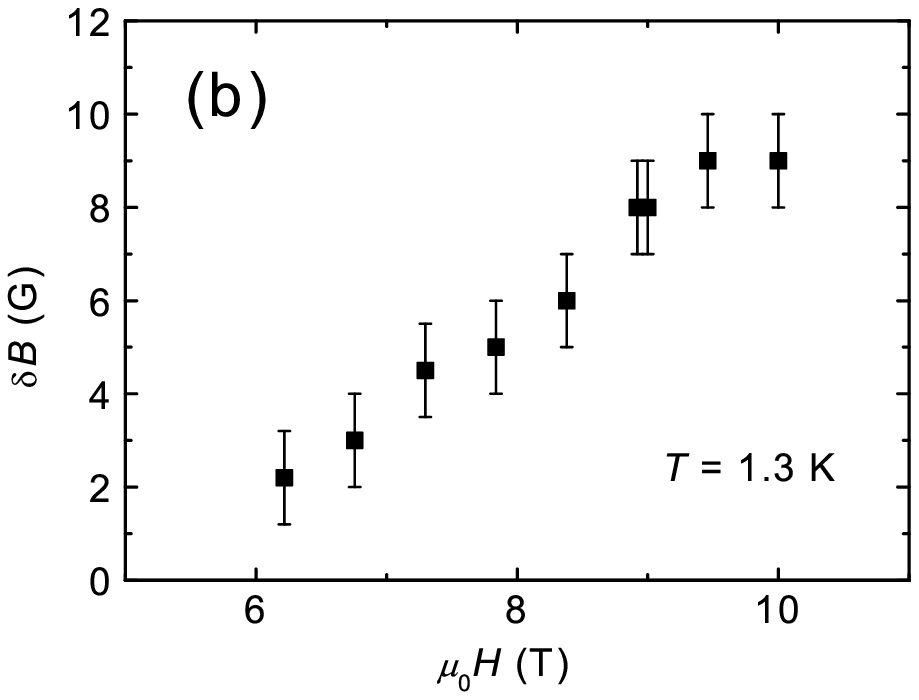}
       \caption{Maximum induction difference $\delta B$ as a function of
       temperature (a) and magnetic field (b).
       Solid line in (a) is calculated from~\cite{GORDON}. Inset shows
       the theoretical phase diagram~\cite{GORDON} with
       dotted lines (a) and (b) indicating the two measuring tracks.}
     \label{f-3}
     \end{center}
     \end{figure}
Fig.~\ref{f-2} shows typical $B(H)$ traces of Hall probes $B_{1}$
and $B_{5}$ of the L-array over three dHvA periods at 10~T and
1.3~K. In each paramagnetic part of the dHvA period two different
inductions are measured at the surface of the sample whereas the
induction is homogeneous in the diamagnetic part. The measured
traces are reversible for increasing and decreasing magnetic field.
We ascribe the measured difference between the induction of
neighboring Hall probes to the existence of Condon domains. The
maximal induction splitting $\delta B$ in a dHvA period was measured
as a function of temperature at 10~T (see Fig.~\ref{f-3}a) and as a
function of field at 1.3~K (see Fig.~\ref{f-3}b). At 10~T, the phase
boundary is crossed at about 3~K. At 1.3 K, the crossing occurs at
about 5~T. The field and temperature range for the occurrence of the
induction splitting is in agreement with the $T$-$H$ phase diagram
for the Condon domain state in Ag, as shown in the inset of
Fig.~\ref{f-3}a for the theoretically calculated phase-diagram of Ag
with Dingle temperature 0.2~K~\cite{GORDON}. The solid line in
Fig.~\ref{f-3}a is the calculated induction splitting $\delta
B$~\cite{GORDON} with the phase transition temperature (3~K) and the
maximum splitting in silver (12~G) measured by Condon and
Walstedt~\cite{CONWAL} as parameters.

\begin{figure}%[!h]
     \begin{center}
    \leavevmode
       \includegraphics[width=0.95\linewidth]{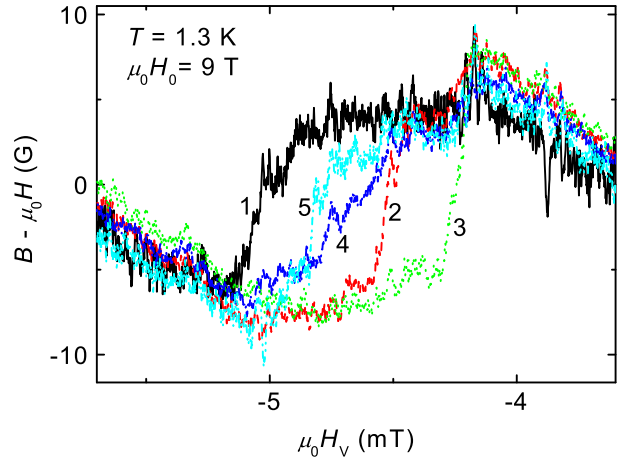}
       \caption{Example of successive transitions for five T-probes between diamagnetic and
       paramagnetic phase. The sweep rate was 0.5~mT/min.
       }
     \label{f-4}
     \end{center}
\end{figure}

\begin{figure}%[!h]
     \begin{center}
    \leavevmode
       \includegraphics[width=0.95\linewidth]{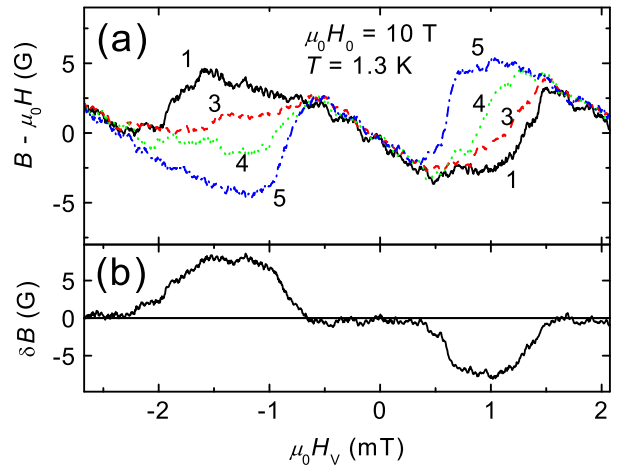}
       \caption{(a) Example of transitions for two dHvA
       periods for L-probes. The transition order is always either 1-3-4-5 or opposite.(b) shows the
       difference $B_{1}-B_{5}$ vs $H$.
       }
     \label{f-5}
     \end{center}
     \end{figure}

\begin{figure}%[!h]
     \begin{center}
    \leavevmode
       \includegraphics[width=0.95\linewidth]{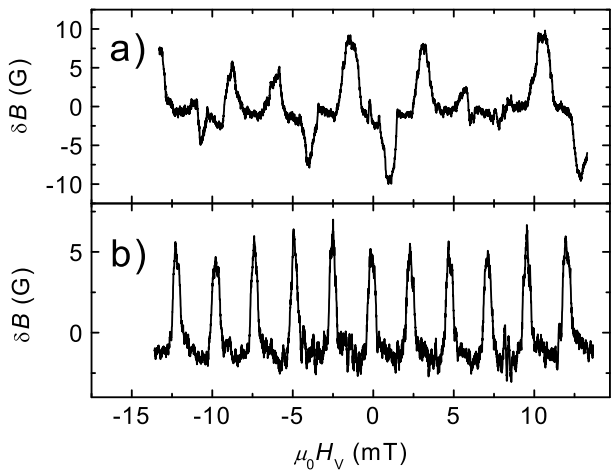}
       \caption{(a) "Pendulum" effect for sample surface oriented
       normal to $\mathbf{H}$ showing a regular change of sign in the  difference signal
       $\delta B = B_{1}-B_{5}$ during 4 or 5 periods.
       (b) Same dependence for the 13$^{\circ}$ tilted sample. In both the temperature was 1.3~K and
       the offset magnetic field was $\mu_0 H_0=10$~T.
       }
     \label{f-6}
     \end{center}
     \end{figure}

An anomalous alternating transition order of the Hall probes between
the diamagnetic and paramagnetic phase is shown in Fig.~\ref{f-2}.
Although reproducible, the observed order depends strongly on the
experimental configuration. A basically different behavior can be
seen between the T-~and L-probes in, respectively, Figs.~\ref{f-4}
and~\ref{f-5}. No regular transition order was observed for
T-probes. Sometimes, they transit in ascending (1-5) or in
descending order, as if the domain laminae were slightly tilted to
the long axis of the sample. Sometimes, as shown in Fig.~\ref{f-4},
a middle T-probe transits the last or the first, as if the laminae
are bent. In contrast, the order of the L-probes is always 1,2,3,4,5
or reversed as it is shown in Fig.~\ref{f-5}a. This implies that the
domain structure is approximately laminar with the laminae mainly
oriented transverse to the long axis of the sample. However, we
found that the L-probe sequence changes alternately between dHvA
periods which implies that the domain-wall movement changes
direction along the long sample axis between successive dHvA
periods. $\delta B=B_{1}-B_{5}$, shown in Fig.~\ref{f-5}b, changes
sign alternately during four or five periods (see Fig.~\ref{f-6}a).
Then the, what we will call, "pendulum" effect breaks down during
two periods where the transition order is not clear. After this the
pendulum effect repeats.

On a slightly tilted sample the pendulum effect disappears
completely. Fig.~\ref{f-6} shows the change of the situation after
tilting the sample. After rotation around the long sample axis by
13$^{\circ}$ domains draw up to a regular laminar structure oriented
always transverse to the long axis. A similar behavior was observed
in white tin in the intermediate state~\cite{SHARVIN,LIVINGSTONE}
indicating the preference of domain walls to align along the sample
surface. Furthermore, the transition order is now the same for all
dHvA periods (see Fig.~\ref{f-6}b). The rotation of the silver
single crystal with respect to the magnetic field affects the dHvA
frequency spectrum. Only one dHvA frequency ("belly"~\cite{SHOEN})
remains for the 13$^{\circ}$ tilted sample. The beating pattern in
the oscillatory dHvA signal of the magnetization for the
perpendicular field orientation might play a role in the occurrence
of the pendulum effect. In this respect we note that the dHvA
frequencies (belly orbit at 47300~T and rosette orbit at 19000~T)
for the perpendicular field orientation would be compatible with the
observed pendulum effect of Fig.~\ref{f-6}a.

The transitions of the individual Hall probes are very sharp
compared to the whole field range of the domain state in neighboring
Hall probes (see Fig.~\ref{f-4}). This means that the thickness of
the domain wall is much smaller than the period of the domain
structure. We have never seen more than one transition of a Hall
probe in a period. This implies that we saw always only one boundary
between para- and diamagnetic phases in an array of 5 successive
Hall probes meaning that the period of the domain structure is
certainly larger than the distance of $\approx150~\mu$m between the
edge L-probes. This under limit for the domain period ($p$) should
be compared with the value obtained from the square-root averaged
expression $p\propto \sqrt{w t}$ for a sample with thickness ($t$)
and domain wall thickness ($w$)~\cite{SHOEN}. With the proposed
cyclotron radius for the wall thickness ($1~\mu$m at 10 T in Ag),
one obtains at least a 5 times smaller value
($\approx30~\mu$m)~\cite{CONWAL}. Therefore, from our experiments we
find a wall thickness of at least $20~\mu$m. This is in agreement
with the observation that two neighboring middle L-probes at a
distance of $40~\mu$m show often intermediate but different
induction values. Therefore, the thickness of a domain wall can not
be much smaller than 20~$\mu$m. As the real domain pattern turns out
to be somewhat bigger than expected, we need either a new set-up
with better adapted Hall-probe distances or a scanning Hall probe
for more detailed measurements of the domain structure.

Exactly the same measurements as presented above were performed on a
beryllium sample cut from the same single crystal where Condon
domain formation was observed using muon spectroscopy~\cite{SOLT}.
The sample was prepared with a surface quality comparable to the Ag
crystal. Even though the expected $\delta B$ inside the crystal is
ten times higher than in silver, we did not find $\delta B>2$~G on
the sample surface. The attempt of Condon and Walstedt to find
domains in beryllium by NMR was not successful,
either~\cite{CONWAL}. The authors gave explanations related to the
quadrupole broadening and the long nuclear thermalization time in
beryllium. However, now we believe that the main reason is the
absence of induction splitting $\delta B$ at the sample surface.
This could be an intrinsic property of beryllium related to its
anisotropic magnetostriction~\cite{LYKOV}.

In conclusion, Condon domains in silver with induction splitting up
to 10~G were observed by micro Hall probes at fields and
temperatures which are in agreement with the theoretically estimated
phase diagram. A laminar domain structure was found with the
orientation mainly transverse to the long sample axis. The domain
transitions are always reversible for increasing and decreasing
magnet field. For a slightly tilted sample the strange pendulum
effect disappears, and the transitions occur in the same order for
all dHvA periods. The domain period is not smaller than 150~$\mu$m
and the domain wall thickness must be about 20~$\mu$m. Condon
domains in beryllium do not emerge to the surface.

We are grateful indebted to M.~Schlenker for his support and
continuous interest to this work, to I.~Sheikin and V.~Mineev for
fruitful discussions, and to J.~Marcus for his help in sample
surface preparation. F.~Schartner is acknowledged for the
preparation of the Hall probes.

\end{document}